\documentclass[aps,prd,final,letterpaper, superscriptaddress]{revtex4}

\usepackage{amssymb,amsmath}
\usepackage{fullpage}
\usepackage{graphicx}
\usepackage{amsmath}		
\usepackage[margin=1.0in]{geometry}
\usepackage{setspace}
\usepackage{color}
\usepackage{fancyhdr}
\usepackage{collcell}
\usepackage{datatool}
\usepackage{environ}
\usepackage{hyperref}

\definecolor{darkgreen}{rgb}{0,0.5,0}

\newcommand{\coupling}{g_{a\bar{\psi}\psi}}

\begin{document}

\title{ Spin Precession Experiments for Light Axionic Dark Matter}

\author{Peter~W.~Graham}
\affiliation{Stanford Institute for Theoretical Physics, Department of Physics, Stanford University, Stanford, CA 94305}
\author{David~E.~Kaplan} 
\affiliation{Department of Physics $\&$ Astronomy, The Johns Hopkins University, Baltimore, MD 21218}
\author{Jeremy~Mardon}
\affiliation{Stanford Institute for Theoretical Physics, Department of Physics, Stanford University, Stanford, CA 94305}
\author{Surjeet~Rajendran}
\affiliation{Berkeley Center for Theoretical Physics, Department of Physics, University of California, Berkeley, CA 94720}
\author{William~A.~Terrano}
\affiliation{Center for Experimental Nuclear Physics and Astrophysics, University of Washington, Seattle, WA 98195-4290}
\affiliation{Physikdepartment, Technische Universit{\"a}t M{\"u}nchen, D-85748 Garching, Germany}
\author{Lutz~Trahms}
\affiliation{Physikalisch-Technische-Bundesanstalt (PTB) Berlin, D-10587 Berlin, Germany}
\author{Thomas~Wilkason}
\affiliation{Department of Physics, Stanford University, Stanford, California 94305}

\begin{abstract}
Axion-like particles are promising candidates to make up the dark matter of the universe, but it is challenging to design experiments that can detect them over their entire allowed mass range.  Dark matter in general, and in particular axion-like particles and hidden photons, can be as light as roughly $10^{-22}$ eV ($\sim 10^{-8}$ Hz), with astrophysical anomalies providing motivation for the lightest masses  (``fuzzy dark matter").  We propose experimental techniques for direct detection of axion-like dark matter in the mass range from roughly $10^{-13}$ eV ($\sim 10^2$ Hz) down to the lowest possible masses.  In this range, these axion-like particles act as a time-oscillating magnetic field coupling only to spin, inducing effects such as a time-oscillating torque and periodic variations in the spin-precession frequency with the frequency and direction of these effects set by the axion field. We describe how these signals can be measured using existing experimental technology, including torsion pendulums, atomic magnetometers, and atom interferometry. These experiments demonstrate a strong discovery capability, with future iterations of these experiments capable of pushing several orders of magnitude past current astrophysical bounds.


\end{abstract}

\maketitle

\tableofcontents
\section{Introduction}

The cold-dark-matter paradigm has been established as a critical part of our understanding of cosmology, but the fundamental nature of this dark matter remain unknown \cite{Clowe:2006eq}.  Axions are among the most well-motivated of the viable dark matter candidates, with many theories of Beyond the Standard Model physics including mechanisms that can produce ubiquitous axions and other ultralight bosons with the correct abundance to match the observed dark matter density \cite{PhysRevLett.38.1440, PhysRevD.16.1791, PhysRevLett.40.223, PhysRevLett.40.279, PhysRevLett.43.103, SHIFMAN1980493, DINE1981199, Dine:1982ah, Preskill:1982cy, Svrcek:2006yi, Arvanitaki:2009fg}. The large parameter space where ultra-light bosons are good dark matter candidates has inspired new interest in experimental searches for axion and axion-like searches \cite{Sikivie:1983ip, Duffy:2006aa, Asztalos:2009yp, Wagner:2010mi, Graham:2011qk, Asztalos:2011bm, Pospelov:2012mt,Stadnik:2013raa, Sikivie:2013laa, PhysRevD.88.035023, PhysRevX.4.021030, Arvanitaki:2014dfa, Stadnik:2014xja, Kahn:2016aff, TheMADMAXWorkingGroup:2016hpc, Abel:2017rtm} as well as other types of ultra-light bosonic dark matter \cite{Jaeckel:2007ch, Povey:2010hs,  Horns:2012jf, An:2013yua, Derevianko:2013oaa, Parker:2013fba, Betz:2013dza, Graham:2014sha, Chaudhuri:2014dla, An:2014twa, Arias:2014ela, Dobrich:2014kda,  Izaguirre:2014cza, Arvanitaki:2014faa, VanTilburg:2015oza, PhysRevD.93.075029, PhysRevLett.115.201301,  PhysRevLett.117.061301, PhysRevA.94.022111, PhysRevLett.114.161301, PhysRevA.93.063630, Arvanitaki:2016fyj, Arvanitaki:2015iga}, with many experiments in progress.


In particular, there has been renewed interest in these types of ultralight dark matter with masses as low as $m_{\rm{a}} \sim 10^{-22}\;\rm{eV}$. This ``Fuzzy Dark Matter'' has a Compton wavelength on the order of the size of dwarf galaxies, which circumvents potential problems associated with structure formation from standard cold dark matter \cite{Khlopov:1985jw, Press:1989id, Sahni:1999qe, Hu:2000ke, Marsh:2013ywa, Schive:2014dra, PhysRevD.95.043541, Lee:2017qve}.
In addition, recent measurements suggest there are slight excesses in the cooling of white dwarfs that could be explained by the addition of ultralight axions \cite{1475-7516-2016-07-036}.  It is interesting to note axions with even lighter masses below $10^{-22} \;\rm{eV}$ can also form a partial contribution to the total dark matter mass density \cite{Hlozek:2014lca, Hlozek:2017zzf}. 
This extreme ultralight dark matter has been the focus of several recent experimental proposals \cite{Arvanitaki:2014faa,  PhysRevLett.117.061301, PhysRevA.94.022111, PhysRevLett.114.161301, PhysRevA.93.063630, VanTilburg:2015oza, PhysRevD.93.075029, Arvanitaki:2016fyj}, but most have focused on scalar dark matter and its couplings. In this paper, we present several experiments that can be modified or created to search for axions at the lightest masses, and evaluate their potential to reach axion couplings several orders of magnitude beyond current astrophysical bounds.

In this paper, we focus exclusively on axion-like particles (which we will refer to simply as axions or short), and focus on the nature and potential discovery of ultralight dark matter composed primarily of these particles. An axion is created as the Goldstone boson of a high-scale symmetry breaking, so we expect it to have a derivative interaction with fermions of the form:
\begin{equation}
\mathcal{L}_{\rm{ax}} = \coupling \partial_{\mu} a\bar{\psi}\gamma^{\mu}\gamma^{5}\psi
\label{eq:spinlag}
\end{equation}
where $\psi$ is the fermion field, $a$ is the axion field, and $g_{a\bar{\psi}\psi}$ is the coupling constant of the axion field to the fermions, which is inversely proportional to the energy scale of the symmetry breaking. If the axion is ultralight (below $\sim 1$ eV), then the phase space density of the dark matter implies there must be many particles per cubic de Broglie wavelength. Much like the large occupation of photons acts to form a coherent electromagnetic wave, we expect that this large number density of axions should act as a classical field, with the particle oscillating around the minimum of its classically quadratic potential with a frequency equal to its mass. The field will then take the form $a(t, \vec{x}) \sim a_0\cos{\left(E_{\rm{a}}t + \vec{p}_{\rm{a}}\cdot \vec{x}\right)}$, where $E_{\rm{a}}$ and $\vec{p}_{\rm{a}}$ are the the energy and momentum, respectively, of the axion. The distribution of energy (and thus frequency) is centered on the mass $m_{\rm{a}}$ of the axion, but since the axions are non-relativistic and moving at a velocity $v$, they have a small spread in their energy set by the kinetic energy of the axion $m_{\rm{a}} v^2$. We can understand the coherent effect of the axion field by analyzing the non-relativistic limit of the above interaction, which gives rise to the following Hamiltonian:

\begin{equation}
H_{\rm{ax}} = -\coupling\vec{\nabla}a \cdot \vec{\sigma}_{\psi} 
\label{eq:spinhamiltonian}
\end{equation}
where $\vec{\sigma}_{\psi}$ is the spin operator of the fermion field $\psi$. This simplification of the Hamiltonian can be understood by recognizing the Lagrangian in Equation \ref{eq:spinlag} as a magnetic dipole moment operator, with a pseudo-magnetic field defined by the gradient (and momentum) of the axion field. This pseudo-magnetic field couples only to spin, and hence all axion-coupled fermions will precess around the direction of the axion dark matter momentum. 

We assume the axion dark matter permeates throughout the galaxy, and so this creates an ``axion wind'' that flows through the Earth at the galactic virial velocity, $\vec{v}\sim10^{-3}\hat{v}$. The momentum of this wind thus can be calculated as $\vec{\nabla}a \sim \vec{p}_a \; a \sim  m_a \vec{v}\; a_0\cos{m_{\rm{a}} t}$. We can further simplify this by recognizing that the energy density of this field can be calculated as $\rho_{\rm{DM}} = \frac{1}{2}m_{\rm{a}}^2a^2$, where $\rho_{\rm{DM}} = (0.04 \;\mathrm{eV})^4$ is the established measured energy density of dark matter in the galaxy. Using these facts, we can further simplify the Hamiltonian as:

\begin{equation}
\begin{split}
H_{\rm{ax}} \sim \coupling a_0 m_{\rm{a}}\left(\vec{v}\cdot \vec{\sigma}_{\psi}\right)\;\cos{m_{\rm{a}} t} \\
\Rightarrow H_{\rm{ax}} \sim \sqrt{2\rho_{\rm{DM}}}\vec{v} \cdot \vec{\sigma}_{\psi} \cos{m_{\rm{a}} t}
\end{split}
\end{equation}

We can see from this final Hamiltonian that this is analogous to the coupling of spin to a pseudo-magnetic field of size $B \sim g_{a\bar{\psi}\psi}v\sqrt{2\rho_{\rm{DM}}}/\gamma_{\psi}$, where $\gamma_{\psi}$ is the gyromagnetic ratio of the fermion. Since we are interested in the spin precession of fundamental fermions and we do not know the direction of the dark matter relative to the Earth, we will take $\vec{v}\cdot\vec{\sigma} = \frac{1}{2}v$ on average (assuming the spins are collinear with the axion field, and ignoring order one factors accounting for the rotation of the Earth).  We then estimate the size of axion-induced energy splitting per spin as:

\begin{equation}
H_{\rm{ax}} \sim 10^{-25} \; \mathrm{eV}\;\left(\frac{\coupling}{10^{-10}\; \mathrm{GeV}^{-1}}\right)\left(\frac{v}{10^{-3}}\right)\left(\sqrt{\frac{2\rho_{\rm{DM}}}{(0.04\; \mathrm{eV})^4}}\right)\cos{m_{\rm{a}} t}
\end{equation}

Although this is a small effect, making use of a large number of coherent spins can greatly increase the potential for direct detection through this channel, as previously discussed for heavier axions with masses between $10^{-14} \;\rm{eV}$ and $10^{-7} \;\rm{eV}\;$\cite{PhysRevD.88.035023, PhysRevX.4.021030}. 


Experiments designed to search for spin-dependent violation of Lorentz invariance (LIV) are a natural starting point for these ultralight axion searches. The LIV signal manifests itself at lowest order through the non-relativistic Hamiltonian $H = b_{\psi}\cdot\sigma_{\psi}$, where $b_{\psi}$ is the overall energy shift that quantifies the size of the local Lorentz violating field \cite{PhysRevD.55.6760, PhysRevD.58.116002}. We can immediately see a correspondence between this energy shift $b_{\psi}$ and the axion coupling $g_{\rm{a}\bar{\psi}\psi}v\sqrt{2\rho_{\rm{DM}}}$, so we expect the LIV signal to be identical to the axion signal in the zero axion mass limit. These same experiments, including spin-polarized torsion pendulums and atomic magnetometers, can then be used with little to no modification to search for slowly-varying axion fields. We will discuss for each experiment how bounds on these measured values can be interpreted in terms of limits on the axion coupling.  


In this paper, we study the experimental effects of axion-like particles at the lowest axion masses, i.e. $m_{\rm{a}} \sim 10^{-14} - 10^{-22}\; \mathrm{eV}$, including the mass range of Fuzzy Dark Matter. In this mass regime, the axion field oscillates at frequencies from $100 \;\mathrm{nHz}$ (roughly an inverse year) to $100 \;\mathrm{Hz}$, and so the axion signal is modulated at experimentally accessible timescales. The experiments consist of many measurements each with an inverse timescale $T_{\rm{meas}}^{-1}$ that is usually within the frequency range of interest in this paper. For the lowest axion masses $m_{\rm{a}} \ll T_{\rm{meas}}^{-1}$, and the axion signal is essentially constant over an individual measurement. In this case, the axion induced precession would vary across measurements, with the variation appearing at the axion Compton frequency. At the other end of the frequency range, where $m_{\rm{a}} \gg T_{\rm{meas}}^{-1}$, the axion signal oscillates many times per measurement and requires techniques with sub-$T_{\rm{meas}}$ resolution to extract.

We will discuss several existing experiments that, with some modifications and optimizations, can be used to search for axion dark matter at the extreme ultralight frontier. We note that while the idea that torsion balances and atomic co-magnetometers can be used to search for axion dark matter has been previously mentioned \cite{Stadnik:2015upa, Flambaum_talk}, here we go through a full consideration of the reach of these experiments, and the shape of our full sensitivity is different than previous claims. In particular, in contrast to previous work, we find that these techniques can be used to search for axion dark matter significantly beyond the astrophysical bounds.  It is also interesting to note that a nucleon spin precession measurement for dark matter has recently been performed \cite{Abel:2017rtm}, which, although it does not reach beyond the astrophysics bounds, does present the first dedicated analysis for this type of coupling in this mass range.

For consistency, we demonstrate projected sensitivities for all experiments after 1 year of integration based on current and potential future experimental parameters, with optimistic assumptions about the possibility of removing backgrounds. In Section \ref{sec:torsion_pend}, we discuss spin-polarized torsion pendulum searches. In Section \ref{sec:magnetometers}, we discuss high-precision atomic magnetometers. In Section \ref{sec:atom_intf}, we propose a new experiment to search for axions based on existing atom interferometry technology. In Section \ref{sec:vectors}, we discuss the possibility to extend the searches described above to vector dark matter. We conclude in Section \ref{sec:conclusions} with a discussion of the landscape for new experiments designed to detect axions.




\section{Torsion Pendulum}
\label{sec:torsion_pend}


The axion signal manifests itself as a spin precessing around the direction of the axion velocity vector. The most classically intuitive observable is a torque on the spin that is proportional to the axion coupling. We can estimate the torque per spin from the Hamiltonian in Equation \ref{eq:spinhamiltonian} as a cross product between the direction of the axion field and the spin of the particle. For a macroscopic object made up of $N_p$ aligned spins, we thus have a total torque on the object of:

\begin{equation}
\vec{\tau}_{\psi} \sim \coupling  N_p \frac{1}{2}\vec{v}\sqrt{2\rho_{\rm{DM}}} \cos{m_{\rm{a}} t}.
\label{signal}
\end{equation}
For typical dark matter parameters and experimental sizes, the size of this signal is roughly
\begin{equation}
\tau \sim 10^{-2} \; \mathrm{eV}\;\left(\frac{\coupling}{10^{-10}\; \mathrm{GeV}^{-1}}\right)\left(\frac{N_p}{10^{23}}\right)\left(\frac{v}{10^{-3}}\right)\left(\sqrt{\frac{\rho_{\rm{DM}}}{(0.04\; \mathrm{eV})^4}}\right)\cos{m_{\rm{a}} t}.
\end{equation}
This effect will thus manifest itself as a small torque on a material with a large spin moment. In this section, we discuss how this effect manifests itself in the spin-polarized torsion balance at the University of Washington, and discuss strategies to search for the axion using this experiment \cite{PhysRevD.78.092006,Adelberger:2009zz, Heckel:2013ina, Terrano:2015}.

\subsection{Torsion Pendulum Searches}
\label{subsec:torsion_pend_bg}

The spin-polarized torsion pendulum at the University of Washington is an ideal instrument to look for the spin-coupled axion, as it contains a large net spin-moment and minimal magnetic moments. To achieve this, the pendulum is made from a clever combination of two magnetic materials that have different spin polarizations at the same magnetization.  The pendulum contains a total of number of polarized electron spins $N_p \sim 10^{23}$  and a magnetic moment of less than $10^{-5} \;\mathrm{J/G}$, making the experiment sensitive to the $g_{aee}$ coupling \cite{PhysRevD.78.092006}.

Similar to the experiment proposed in \cite{PhysRevD.93.075029} to search for ultra-low mass scalar and vector dark matter, a torque due to the oscillating axion field will appear at 6 different frequencies corresponding to the combination of the axion frequency $\nu_{\rm{a}} = m_{\rm{a}}/2\pi$, the turntable frequency $\nu_{\rm{TT}}$, and the Earth rotation frequency $\nu_{\oplus}$
\begin{gather}
\label{eq:DM-twist}
\nu_{\rm{sig}} = \nu_{a} \pm \nu_{\rm{TT}} + \Big\{ \begin{smallmatrix} +\nu_{\rm \oplus} \\ 0 \\ -\nu_{\rm \oplus} \end{smallmatrix} \Big\} \, ,
\end{gather}
%

The coefficients of this signal are given in Equation \ref{signal}. The axion signal appears at frequencies set by fundamental physics, allowing it to be more easily distinguished from systematic backgrounds like gravity gradients, magnetic field drifts, and turntable tilt \cite{PhysRevD.78.092006}. Since the torsion pendulum is typically not limited by the read-out system, a broadband search over the axion dark matter frequency space is most appropriate.

The fundamental noise limit from the experiment comes from two primary contributions: thermal noise from the fiber (which dominates at low axion frequencies) and angular readout noise from the laser readout system (which dominates at high frequencies). The thermal noise power for the fiber takes the form:

\begin{equation}
\mathcal{P}_{\rm{th}} = \frac{4 T \kappa}{2\pi \nu_{\rm{sig}} Q}
\end{equation}
where $T$ is the temperature of the fiber, $\kappa$ is the fiber torsion constant, and $Q$ is the quality factor of the fiber. Rotating the balance places $\nu_{\rm{sig}} \sim \nu_{\rm{TT}}$ for all axion frequencies below $\nu_{\rm{TT}}$, greatly reducing the thermal noise in searches for low-frequency signals. For existing balances we have typical thermal noise levels of:
\begin{equation}
\sqrt{\mathcal{P}_{\rm{th}}} \sim 4\times10^{4}\;\frac{\mathrm{eV}}{\sqrt{\mathrm{Hz}}}\times\sqrt{\left(\frac{T}{300 \;\mathrm{K}} \right)\left(\frac{\kappa}{0.185 \;\mathrm{erg/rad}}\right)\left(\frac{1 \;\mathrm{mHz}}{\nu_{\rm{sig}}}\right)\left(\frac{2000}{Q}\right)}
\end{equation}
At a turntable frequency of $\nu_{\rm{TT}} \sim 1\;\mathrm{mHz}$, this corresponds to a thermal noise of $\sqrt{\mathcal{P}_{\rm{th}}} \simeq 10^{-15} \;\mathrm{N\cdot m}/\sqrt{\rm{Hz}}$. In the case of $\nu_{\rm{a}} > \nu_{\rm{TT}}$, we have $\nu_{\rm{sig}} \sim \nu_{\rm{a}}$, and so the thermal fiber noise is even lower at this higher frequency. The thermal noise can be reduced by rotating the table at higher frequencies, reducing the temperature of the system, or using lower-noise fibers that have smaller $\kappa$  or larger $Q$. 

The readout noise takes the form:
\begin{equation}
 \;\;\; \mathcal{P}_{\tau} = I^2\left(2\pi\right)^4\left((\nu_{\rm{sig}}^2 - \nu_0^2)^2 + \frac{\nu_0^4}{Q^2}\right)\mathcal{P}_{\theta}
\label{eq:readout}
\end{equation}
where $I$ is the moment of inertia of the pendulum, $\nu_0$ is the resonance frequency of the apparatus, and $\mathcal{P}_{\theta}$ is the noise power for the angular readout. Using existing experimental parameters for the angular readout system and assuming the frequency of the axion dominates over the frequency of the pendulum, we can also estimate the size of the readout noise \cite{PhysRevD.78.092006}

\begin{equation}
\sqrt{\mathcal{P}_{\tau}} \sim 2\times10^5 \;\frac{\mathrm{eV}}{\sqrt{\mathrm{Hz}}}\times\left(\frac{I}{10^{-4}\;\mathrm{kg}\cdot\mathrm{m}^2}\right)\left(\frac{\nu_{\rm{sig}}}{1 \;\mathrm{Hz}}\right)^2\left(\frac{\sqrt{P_{\theta}}}{10^{-9}\;\mathrm{rad}/\sqrt{\mathrm{Hz}}}\right)
\end{equation}

The noise from the angle-readout increases quadratically with frequency above $\nu_0$, and so rapidly becomes the leading noise source above the resonance frequency of the pendulum. To optimize the two noise sources over the total frequency range, the turntable frequency should be located where the thermal noise and angular readout noise are equal, but in practice it is often slightly lower than that due to experimental constraints. The turntable frequency should be lower than or near the resonance frequency of the pendulum to suppress unwanted rotational modes due to imperfections in the turntable.

Due to these constraints, it is important to note how the noise scales with the size of the torsion pendulum. Assuming the density of the material stays constant, both the signal and the thermal noise are proportional to the total mass of the pendulum, since the torsion constant of a fiber scales with the square of the mass it can support. Since the readout noise scales with the pendulum moment of inertia, smaller pendulums will have improved signal to noise at high frequencies.


In this analysis we assume that the torsion pendulum will remain the same size for all future iteration, with $I \simeq 10^{-4} \;\rm{kg\cdot m^2}$, $\kappa = 0.185 \;\rm{erg/rad}$, and resonance frequency of $\nu_{0} \simeq 5 \;\rm{mHz}$. Current experiments take $\nu_{\rm{TT}} \simeq \nu_{0}/10$, but for future experiments, this can be pushed up closer to $\nu_{\rm{TT}} \simeq \nu_0$ \cite{PhysRevD.78.092006}. 


\subsection{Sensitivity Estimate}

We can estimate the strength of the axion field coupling we are sensitive to in these experiments by knowing the total integration time and how the noise scales with time. We use the noise scaling as derived in \cite{PhysRevD.88.035023}, which takes into account the coherence time of the axion dark matter $\tau \simeq 10^6/m_{\rm{a}}$, and a total integration time of $T_{\rm{int}} \sim 1$ year. 

Combining the signal estimate in Equation \ref{signal} with the background estimate in section \ref{subsec:torsion_pend_bg}, we find the exclusion sensitivity of the torsion pendulum to the axion assuming a signal-to-noise ratio (SNR) $\sim 1$. Since there is no a priori theoretical guess of the axion mass, the detection sensitivity (which requires much higher SNR) is further reduced by the large number of frequencies that must be tested. To estimate the exclusion sensitivity, we analyze these noise sources for the current experimental parameters (Current), near-term improved parameters (Upgrade), and some optimistic future parameters involving a cryogenic torsion balance (Future). These parameters are described in Table \ref{parameter_table}. Note that the result from the current analysis is similar to the limit on DC LIV as measured in \cite{PhysRevD.78.092006}, which bounded the electron coupling to the Lorentz violating field of $b_{\rm{e}} < 3.4 \times 10^{-22} \;\rm{eV}$. This corresponds to a coupling of $g_{aee} \simeq 1.5 \times 10^{-7} \;\rm{GeV}^{-1}$, which differs slightly from the plot due to our optimistic assumptions about integration time.

\begin{table}
\begin{center}
 \begin{tabular}{|c || c | c | c | c |} 
 \hline
  & Current & Upgrade & Future\\
 \hline\hline
 $T$ (K) & 300 & 300 & 6 \\ 
 \hline
 $Q$ & 2000 & $10^6$ & $10^8$\\
 \hline
 $\sqrt{\mathcal{P}_{\theta}}$ (rad/$\sqrt{\mathrm{Hz}}$) & $10^{-9}$ & $10^{-12}$ & $10^{-18}$ \\ 
 \hline
 $\nu_{\rm{TT}}$ (mHz) & $0.5$ & $0.5$ & $5$ \\
 \hline
\end{tabular}
	\caption{Parameter estimates for various runs of the torsion pendulum. ``Current'' estimates are for the balance at the University of Washington used in the measurements resported in \cite{PhysRevD.78.092006}. ``Upgrade'' estimates are with the same pendulum, but with upgrades the group is planning for the next few years. ``Future'' estimates push the physical limits of what may be possible with a torsion pendulum technique, assuming several improvements that have not yet been experimentally demonstrated.}
\label{parameter_table}
\end{center}
\end{table}

\begin{figure*}
\includegraphics[width = 0.8\textwidth]{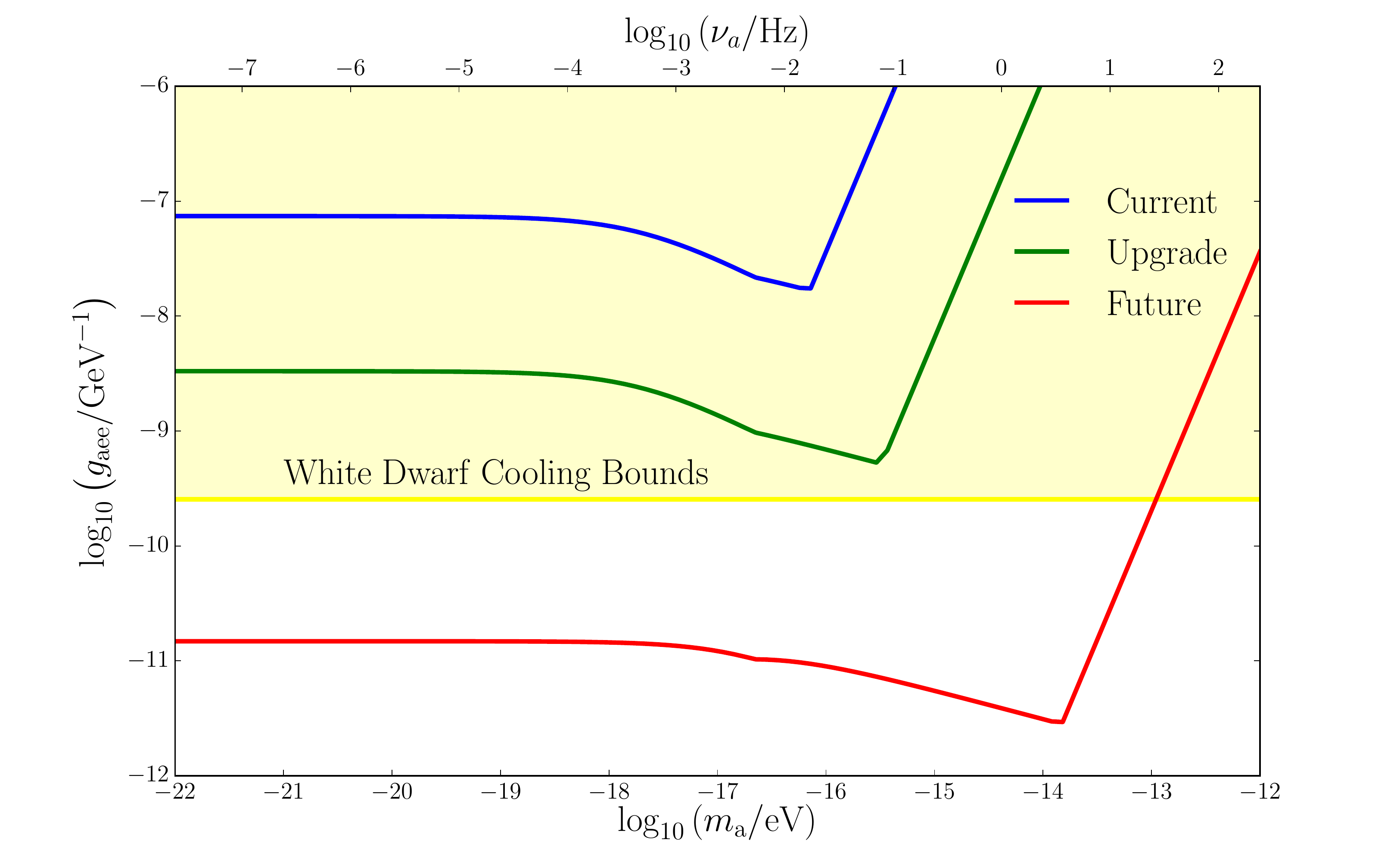}
\caption{Estimate of the exclusion sensitivity reach of the spin-polarized torsion pendulum for the $g_{aee}$ coupling over an integration time of 1 year. The different iterations of the experiment have improved thermal and readout noise parameters as given in Table \ref{parameter_table}. White dwarf and red giant cooling observations currently favor excess cooling in the $\sim 10^{-9}-10^{-10}$ coupling strength range, making this a particularly interesting target range for experiments \cite{1475-7516-2016-07-036}. We have the expected behavior for the lowest frequencies due to the flat frequency response of the fiber noise. For higher frequencies, the sensitivity begins to fall off slowly due to coherence time of the axion, and then more quickly once the sensitivity is dominated by the angle read-out noise.  All curves are expected sensitivity curves, and in particular the ``current'' line is not currently a limit, it is only a projection for where current technology could reach given the assumptions laid out in the text.}
\label{pend_sens_est}
\end{figure*}

We plot this sensitivity over a range of masses in Figure \ref{pend_sens_est} and compare it to the astrophysical bounds on the axion coupling from White Dwarf cooling \cite{Raffelt:2006cw}. The sensitivity can be understood in two different regimes separated by the turntable frequency. We see that the experiment is particularly sensitive to low mass axions, including down to the lower mass limit of $10^{-22}$ eV. In this region, the ultralight axion appears almost as a DC effect, so the noise is dominated by the thermal fiber noise. We can see from Equation \ref{eq:DM-twist} that because $\nu_{a} < \nu_{\rm{TT}}$, the signal appears at sidebands of the primary turntable frequency. The noise of the system is set by the fiber thermal noise at $\nu_{\rm{TT}}$, leading to a sensitivity that is independent of the axion mass. When $\nu_{\rm{a}} > \nu_{\rm{TT}}$, the noise is set by the fiber thermal noise at $\nu_{\rm{a}}$, which improves with increasing frequency. In this range the ability to operate with $\nu_{\rm{TT}} > \nu_{0}$ would improve sensitivity across the entire low-frequency range. Once $\nu_{\rm{a}} > \nu_{0}$, the signal oscillates more rapidly than the pendulum can respond, and so the measurement of the pendulum twist angle limits the sensitivity of the experiment (as shown in Equation \ref{eq:readout}. 


For masses $m_{\rm{a}} > Q/T_{\rm{int}}$, we are limited by the finite coherence time of the axion dark matter. The total signal sensitivity thus decreases with the quarter root of the total integration time. For our particular integration time, this transition occurs at $m_{\rm{a}} \sim 10^{-17} \; \rm{eV}$, as indicated by the corresponding kink in Figure \ref{pend_sens_est}. For higher masses, the sensitivity becomes dominated by the readout noise, so the noise scales with frequency like $\nu^{-2.25}$ (from the combined effect of the time scaling with the readout noise, as demonstrated by the second kink around $10^{-14}\; \rm{eV}$ for the bottom curve). The experiment quickly loses sensitivity to higher masses. However, this mass range is covered by other experiments \cite{PhysRevX.4.021030}, so there is only a small amount of axion parameter space that is not covered (this space should be completely covered if optimistic projections for torsion pendulum improvements can be met).

We find that current experimental parameters do not quite reach the astrophysical limits, but that future experiments could allow for significant reach beyond these bounds. The future of torsion pendulum experiments allows for significant improvement in technology beyond optimization of the different thermal/readout noise parameters. With the experimental parameters considered here, the optimal turntable frequency is quite far above the resonance frequency of the pendulum, so the low-frequency noise can be further reduced through optimization of the pendulum size. Furthermore, a potential improvement in signal gain could be to use superconducting magnets on the torsion pendulum rather than rare-earth magnets, as this would increase the total number of spins/mass by up to a factor of 5.

\section{Atomic Magnetometers}
\label{sec:magnetometers}


The Larmor precession of spin-coupled axion dark matter can also be directly observed using high precision magnetometers. High-precision atomic magnetometers have emerged within the past few years as powerful alternatives to conventional SQUID magnetometers, reaching comparable sensitivities in the range of $1 \,\rm{fT}/\sqrt{\rm{Hz}}$. Since atomic magnetometers involve the precise measurement of atomic spins, they can also probe new fields that couple directly to spin. To accurately separate the anomalous signal from the magnetic effects, these experiments typically measure the precession of two different species of atoms with different spin in the same volume.  These sensitive co-magnetometers are well-suited for searches for axion dark matter that couples to nuclear spin. 

For our analysis, we consider two different magnetometers that were used to search for anomalous couplings to spin associated with Lorentz invariance violation (LIV), a search similar to a search for low-frequency spin-coupled axion dark matter \cite{PhysRevLett.105.151604, PhysRevD.82.111901}:
\begin{itemize}
\item Electron-nucleon co-magnetometer at Princeton University
\item Nucleon-nucleon co-magnetometer at Physikalisch-Technische Bundesanstalt 
\end{itemize}
We will review both experiments and their relative similarities and differences, and propose analysis strategies for searching for low-mass axions using the currently existing setups. For both experiments, we also re-interpret the bounds placed on static LIV parameters as bounds on the axion coupling at the lowest frequencies, with an upper frequency bound set by the inverse integration time of the experimental shots. These bounds are calculated specifically for both experiments and are plotted alongside projected axion sensitivities given by noise limits and a total integration time of 1 year.

\subsection{Electron-Nucleon Co-Magnetometer}
\label{subsec:romalis_mag}

The experiment functions as a self-compensating co-magnetometer by exploiting the interaction between the electron spins in K and the nuclear spins in $^{3}$He. The densities and polarization of the two species are arranged such that small changes in the background magnetic field are canceled. However, the electron and neutron spins will have different couplings to a new spin interaction, and so the influence of the new spin interactions will not cancel in the co-magnetometer. This produces a non-zero atomic transverse polarization which can then be read-out through measuring the optical rotation of a polarized probe laser traversing the cell. This produces a relative polarization of \cite{PhysRevLett.105.151604}:

\begin{equation}
P_x = P^e_z\gamma_e T_2\left(\beta_e - \beta_N\right)
\end{equation}
where $P^e_z$ is the average polarization of the electrons in the cell (set to be 0.5 for optimal magnetometer performance), $T_2$ is the spin relaxation time, and $\beta$ is the ``effective'' magnetic field of the anomalous interaction. This transverse polarization corresponds to the total phase accrued by a precessing spin with Larmor frequency $\gamma_e \beta$ after some spin relaxation time $T_2$ (roughly 3 ms for K). n this form, we can interpret the effective magnetic field from the relation $\gamma_{\psi}\beta = g_{a\psi\psi}v\sqrt{2\rho_{\rm{DM}}}$, so we can see that we have two contributions to this ``effective'' magnetic field, one from the electron axion coupling, and another from the nuclear axion coupling.

Since the axion field is oscillating, the signal size will be the total phase due to the axion during the relaxation time of the atoms.The total transverse polarization will then be: 

\begin{equation}
P_x = P^e_z\gamma_e \left(\frac{g_{aee}}{\gamma_e} - \frac{g_{\rm{aNN}}}{\gamma_N}\right)\frac{v\sqrt{2\rho_{\rm{DM}}}}{m_{\rm{a}}}\sin{m_{\rm{a}} T_2}
\label{eq:osc_pol}
\end{equation}
where $\gamma_e, \gamma_N$ are the electron and neutron gyromagnetic ratios, respectively. Equation \ref{eq:osc_pol} explicitly shows how this experiment is sensitive to both the electron and nucleon coupling. If we assume that $g_{aee} \sim g_{\rm{aNN}}$, the effect from the axion-electron coupling is $10^3$ times smaller than the effect from the axion-nuclear coupling. Thus, we treat $g_{\rm{aNN}}$ as the primary measurement for this experiment, although in principle this experiment is also sensitive to $g_{aee}$ with reduced sensitivity.

For $m_{\rm{a}} \ll T_2^{-1}$, the magnetometer sensitivity will approximate the steady state DC response of the magnetometer, and so previous searches for LIV using K-$^3$He co-magnetometers gives us an estimate of the sensitivity of this technique for ultralight axions. For $m_{\rm{a}} \gg T_2^{-1}$, the dark-matter induced precession changes directions multiple times within a single relaxation time, and so the experiment is only sensitive to the amplitude of the oscillation, which decreases as $m_{\rm{a}}^{-1}$. 

Using typical experimental and theoretical parameters, we can estimate the size of the amplitude of these oscillations. To show the relevant features of both regimes, we estimate the size of the signal at the dark matter mass $m_a \simeq T_2^{-1} \simeq 2\pi \times 50 \;\rm{Hz}$:

\begin{equation}
\begin{gathered}
P_x \simeq 2 \times 10^{-9} \left(\frac{P^e_z}{0.5}\right)\left(\frac{\gamma_e}{1.2\times10^{-8}\;\mathrm{eV/T}}\right)\left(\frac{1.3\times10^{-11}\;\mathrm{eV/T}}{\mu_{\rm{N}}}\right)\\
\times\left(\frac{g_{\rm{aNN}}}{10^{-10}\;\mathrm{GeV}^{-1}}\right)\left(\frac{v}{10^{-3}}\right)
\sqrt{\frac{\rho_{\rm{DM}}}{(0.04\; \mathrm{eV})^4}} \times \left(\frac{2\pi \times 50 \;\mathrm{Hz}}{m_{\rm{a}}}\right)
\label{eq:pol_signal}
\end{gathered}
\end{equation}
Using Equation \ref{eq:pol_signal}, we can easily estimate both the maximum signal size of the co-magnetometer as well as how it decreases for higher axion masses.

The sensitivity of the co-magnetometer as of 2011 is listed as $\delta B \sim 1 \;\mathrm{fT}/\sqrt{\mathrm{Hz}}$, which corresponds to a polarization sensitivity of $\delta P_x \sim 10^{-7} \;1/\sqrt{\mathrm{Hz}}$. This sensitivity is largely limited by various sources of technical noise.  The shot-noise limited magnetic field sensitivity is quoted as $\delta P_x = \sqrt{T_2/N}$, where $N$ is the total number of atoms in the volume of the cell (in this experiment, $1.3 \times 10^{16}$ K atoms and $1.9\times 10^{22}$ He atoms. In an ideal experiment running at the shot noise limit, this would allow phase sensitivity of $\delta P_x \sim 5 \times 10^{-10} \; 1/\sqrt{\mathrm{Hz}}$, a factor of nearly 200 improvement over the current limitation. The noise limit imposed by the projection of the transverse magnetization as discussed in \cite{PhysRevX.4.021030} is about an order of magnitude or so below this shot noise limit, and has the same scaling with atomic density as the shot noise so should not be a limiting factor.


The greatest systematic effect limiting the measurement is the gyroscopic pickup of the rotation of the Earth. Due to the fact that the atoms are in an inertial reference frame, they experience a torque of $\tau = L\nu_{\oplus} = \mu \beta$, so they experience an effective magnetic field of $\beta = \nu_{\oplus}/\mu$ at a frequency of $\nu_{\oplus}$. However, the measurement of axion is helped by the fact that in frequency space, this effect is located at a specific frequency set by fundamental physics, which naturally separates it from other laboratory frequencies. As long as the overall background noise can be controlled below the shot noise level, these sideband frequencies should be distinguishable from the gyroscopic systematic effect. Further discussion of this point is in Section \ref{sec:backgrounds}.

\begin{figure*}
\includegraphics[width = 0.75\textwidth]{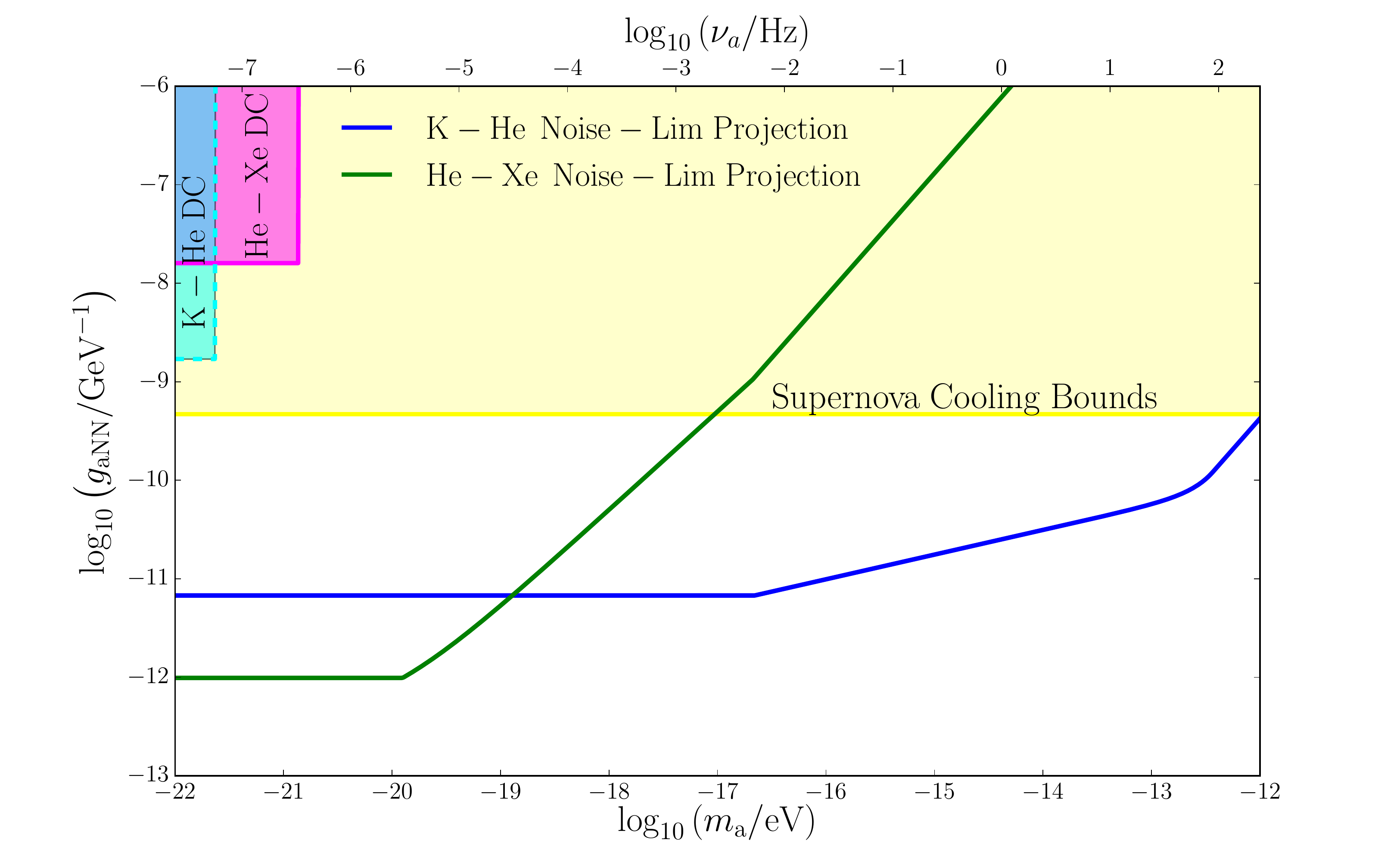}
\caption{Estimate of the exclusion sensitivity reach of the two different magnetometers to the $g_{\rm{aNN}}$ coupling over an integration time of 1 year assuming phase-noise limitation. Current bounds from the DC LIV analyses are re-interpreted as bounds on this coupling for frequencies below the total inverse integration time of the experimental shots \cite{PhysRevLett.105.151604, PhysRevD.82.111901}. For the projected sensitivities, we have the expected behavior for the lowest frequencies, where this effect acts like the DC LIV signal. For higher frequencies, the sensitivity begins to fall off slowly due to coherence time of the axion dark matter and the nuclear spin precession. Note that for the projected sensitivities we have assumed that systematic uncertainties can be overcome and that these experiments can run at their noise limit. Further improvements will come to both experiments by improving relaxation times and densities of the particles involved in the experiment.}
\label{fig:mag_sensitivity}
\end{figure*}

We can interpret the DC LIV limit measured in \cite{PhysRevLett.105.151604} as a limit on a low-mass axion coupling. As described, this experiment ran for two separate runs, each of roughly one month periods. Any effect of an axion with a period greater than this integration time has an identical signature to a DC LIV violating effect. The measured bound of $b_{\rm{N}} < 3.7 \times 10^{-33}\; \rm{GeV}$ can thus be interpreted as a limit on a low-mass axion coupling of $g_{\rm{aNN}} \lesssim 1.6\times10^{-9} \;\mathrm{GeV}^{-1}$. This low-frequency limit is plotted in Figure \ref{fig:mag_sensitivity}. This result can still be used to establish bounds on axion masses of higher frequencies, but we will leave the details of these bounds to a dedicated experimental analysis. 

Using the theoretical shot noise limit described above, we can make an optimistic projection of the sensitivity to the axion coupling $g_{\rm{aNN}}$ shown in Figure \ref{fig:mag_sensitivity}, where we take a total integration time of 1 year and assume optimistic background removal techniques. The total reach of the nuclear coupling should reach past the astrophysics bounds set by SN1987A cooling \cite{Raffelt:2006cw} by almost two orders of magnitude, providing the best bound on this coupling. With further experimental improvements to allow the magnetometer to run at shot-noise limited sensitivity, we can see from the projected line in Figure \ref{fig:mag_sensitivity} that this bound can be improved almost down to the level of $f_a \sim 10^{11}$ GeV, demonstrating strong sensitivity to high energy effects. 

\subsection{Nucleon-Nucleon Co-Magnetometer}
\label{subsec:ptb_mag}

The $^{3}$He --- $^{129}$Xe Co-Magnetometer has a substantially different design to the co-magnetometer described above, but has similar sensitivity to LIV measurements. This co-magnetometer functions by having co-located samples of $^3$He and $^{129}$Xe vapor in a cell whose nuclear spins are polarized along a holding field. The nuclear spins are then tipped perpendicular to and precess around the holding field, and the oscillating magnetic field generated by the precessing nuclear magnetic moments is then measured using a SQUID, with $\sim 1 \mathrm{fT}/\sqrt{\mathrm{Hz}}$ sensitivity. The strength of the guiding magnetic field is chosen such that the two samples have precession frequencies on the order of 10 Hz, which is low enough to allow for long interrogation times but still above the low-frequency vibrational noise picked up by the SQUID detector \cite{PhysRevD.82.111901}. 

To remove the effects of magnetic field drifts in the laboratory, a weighted frequency difference is calculated as $\Delta \nu = \nu_{He} - \frac{\gamma_{He}}{\gamma_{Xe}}\nu_{Xe}$. With perfect compensation, this weighted frequency difference should cancel all magnetic field drifts in the laboratory, leaving a null signal (except for systematic effects, e.g. chemical shift). This allows for a precise measurement of anomalous spin-couplings which would not cancel in this weighted difference. This has been used to search for LIV couplings by measuring sidereal variations in $\Delta \nu$ \cite{PhysRevD.82.111901}. We will discuss two separate methods for analyzing an oscillating signal at low and high frequencies, but these two methods naturally merge into each other and thus are only represented together in Figure \ref{fig:mag_sensitivity}. 

\subsubsection{Low-Frequency Analysis}
%
%

Like in the previous experiment described above, a static anomalous spin-coupling term caused by a DC LIV term has been the subject of an analysis using this co-magnetometer. Due to the rotation of the lab about the earth axis during a sidereal day, variations of the weighted frequency difference $\Delta \nu$ with the Earth rotation frequency $\nu_{\oplus}$ are expected for a static offset as a result of LIV. The total strength of the LIV term can then be found by fitting $\Delta \nu$ to the Earth's rotation frequency and measuring the amplitude of the oscillation. 

An ultra-low mass axion dark matter interaction would cause the weighted frequency difference $\Delta \nu$ to oscillate due to the combined effects of the laboratory rotating relative to the axion field and the oscillation of the axion field itself.  To search for this oscillation, one would measure $\Delta \nu$ as a function of time and fit $\Delta \nu(t)$ at the combined frequencies $\nu_{\oplus}$ and $\nu_{\rm{a}}$. For axions with low frequencies $\nu_{\rm{a}} < \nu_{\oplus}$, we can expect to perform a similar analysis of measuring the amplitude of oscillations, but now with a long term search for the combined Earth and axion frequency. 


The result of the LIV study previously performed was an overall bound on the nucleon coupling to the Lorentz violating field of $b^{\rm{N}}_{\perp} < 3.7 \times 10^{-32} \rm{GeV}$. Like before, this bound can be directly transferred to axion-nuclear interactions, which gives us a bound on the coupling of $g_{\rm{aNN}} < 1.6 \times 10^{-8} \;\rm{GeV}^{-1}$, for frequencies below the inverse total integration time of the experiment of $130$h. This is shown as a shaded bound in Figure \ref{fig:mag_sensitivity}. More recently, this experiment has published a stricter bound on this coupling \cite{PhysRevLett.112.110801}, but these results are currently still under review. This measurement was limited by systematic drifts in the weighted precession frequency, but for future iterations, we assume that these systematics can be overcome and we can measure at the noise limit associated with the precession frequencies.

We note that this is not necessarily a hard limit on the coupling strength. A stronger coupling could have been missed for several reasons, including that at the time when the experiments were made the axion mass induced oscillation might have been in a phase of low amplitude, or the angle between axion wind and quantization axis of the spins might have been unfavorable. For determining a reliable upper limit for the coupling, a series of measurements must be made, at various magnetic field orientations and at various times during the year. Using the noise limits of the co-magnetometer presented in the LIV analysis, we can make a projected limit on the low-frequency bound of this co-magnetometer using a total integration time of 1 year, as shown in Figure \ref{fig:mag_sensitivity}.

\subsubsection{High-Frequency Analysis}

For higher frequencies, the analysis to search for these axions must take a different approach, where the axion-nucleus interaction is identified by the presence of side bands around the Larmor precession frequency. In the SILFIA (``Side Bands in Larmor Frequency Induced by Axions'') detection concept, the co-magnetometer is composed of one nuclear spin and the SQUID. With SILFIA, we can make use of the long relaxation time of a single nucleus like He of up to 100h to obtain a good signal to noise ratio when searching for variations in the frequency of the oscillating transverse magnetization. Here we will present an overview, with more details presented in a separate publication \cite{TrahmsUpcoming}.

Typically, the SQUID will measure an oscillating magnetization due to the precession of the atoms as $B(t) = B_{\rm{T}} \cos \left(2\pi \nu_{\rm{He}}t\right)$, where $B_{T}$ is the field generated by the transverse magnetization of the precessing spins, set by the number of spins in the cell. In practice this amplitude $B_{\rm{T}}$ decays by $e^{-1}$ over the course of the relaxation time, which causes a broadening of the frequency. For simplicity, we will ignore this effect in this analysis. The oscillation of the precession frequency due to the presence of the axion dark matter will generate a SQUID signal of the form

\begin{equation}
B(t) = B_{\rm{T}}\cos{\left(2\pi \nu_{\rm{He}}t + \frac{1}{2}\int_0^t g_{\rm{aNN}}v\sqrt{2\rho_{\rm{DM}}}\sin{m_{\rm{a}} t'} \;dt'\right)}
\end{equation}

This is a frequency modulation of the Larmor precession signal which in the frequency domain, results in a peak of amplitude $B_{\rm{T}}$ at $\nu_{\rm{He}}$, and two first-order side bands with amplitude $B_{\rm{T}}  g_{\rm{aNN}} v\sqrt{2\rho_{\rm{DM}}}/2m_{\rm{a}}$ at frequencies $\nu_{\rm{He}}  \pm m_{\rm{a}}/2\pi$. In the time domain, the signal can be approximated by the superposition of three oscillations

\begin{equation}
B(t) \simeq B_{\rm{T}}\left[\cos{(2\pi \nu_{\rm{He}}t)} + \frac{g_{\rm{aNN}}v\sqrt{2\rho_{\rm{DM}}}}{2m_{\rm{a}}}\bigg( \cos{\left[\left(2\pi\nu_{\rm{He}} + m_{\rm{a}}\right)t + \phi_+\right]} + \cos{\left[\left(2\pi\nu_{\rm{He}} - m_{\rm{a}}\right)t + \phi_-\right]}\bigg)\right]
\label{eq:ptb_sidebands}
\end{equation}

In Equation \ref{eq:ptb_sidebands}, sidebands of higher order are ignored, so that this presentation resembles the result of an amplitude modulation. As long as these sidebands are above the noise floor, $g_{\rm{aNN}}$ can be measured by comparing the size of the side band peak to the Larmor peak. When $m_{\rm{a}}/2\pi < \nu_{\rm{He}}$, then the sidebands should appear as two distinct lines around the precession frequency. For $m_{\rm{a}}/2\pi > \nu_{\rm{He}}$, then we expect to see the axion signal appear as two split lines separated by the twice the precession frequency and centered around the axion frequency. As shown in Equation \ref{eq:ptb_sidebands}, we expect this signal to fall off like $m_{\rm{a}}^{-1}$, as expected from merging with the low-frequency analysis. A most important feature of this detection scheme is that artifact sidebands that are generated by some magnetic interference can be identified in the spectrum of the SQUID signal by the presence of a peak at the frequency of the interference. 

The amplitude $B_{\rm{T}}$ is determined by the observed nuclear magnetic moment and the number of polarized nuclei in the sample. The best typical values for this amplitude are on the order of $B_{\rm{T}} \sim 100 \;\rm{pT}$. For this analysis, we take the noise floor from the SQUID at $\delta B \sim 1 \,\rm{fT}/\sqrt{\rm{Hz}}$. The total signal time set by the life time of transverse spin polarization or the coherence time of axion interaction,  $T = \rm{min}(T_2, \frac{m_{\rm{a}}}{Q})$. 

The requirement for these sidebands to be sufficiently resolved is that the sidebands are well separated from the much higher primary peak, i.e. by several times the linewidth. In a perfect scenario, the carrier peak has a Fourier-limited width of $\pi/T_2$, where $T_2 = 100$h is the longest observed relaxation time of $^3$He. In practice, however, the carrier peak is limited by the temporal stability time of the static magnetic field over the measurement time, which limits the linewidth to a few mHz. This can be overcome by improving the magnetic field control down to the Fourier limit. This limits the effective range of the side-band search to axions of frequency $\nu_{\rm{a}} > 10$ mHz, below which the low-frequency analysis naturally takes over.

\subsubsection{Sensitivity Estimate}

The potential sensitivity of the techniques described above is shown in Figure \ref{fig:mag_sensitivity}. For this analysis we have assumed more modern numbers for the transverse magnetization amplitude and for the SQUID noise than in \cite{PhysRevD.82.111901} improving on $\delta B$ by a factor of three and $B_{\rm{T}}$ by a factor of two, as well as assumed that the systematic uncertainties associated with previous LIV measurements can be overcome and this experiment is only noise-limited. This explains the large increase in sensitivity from the 2010 LIV measurement and the projected measurement. 

Like in the previous experiments, we assume a total integration time of $T_{\rm{int}}$ = 1 year. For axion coherence times less than this integration time, we are able to add the successive shots coherently since we can keep track of the sidereal phase of the recordings. We thus assume we can run individual shots up to the relaxation time $T_2$, which we take to be the limiting Xenon coherence time of $T_2 = 8\rm{h}$, and then run $N = T_{\rm{int}}/T_2$ shots, allowing the signal to build as $\sqrt{N}$. As expected, we find that the method is flat at the lowest frequencies below $T_2^{-1}$, where the low-frequency axion mimics the DC behavior, and then begins rising like $m_{\rm{a}}^{-1}$. There is a second kink in the sensitivity around $10^{-17} \;\rm{eV}$, where the coherence time of the axion is less than the total integration time, causing the signal to fall off by an additional quarter root. The overlap between the low-frequency and high-frequency analysis is expected given that the two analyses are effectively a time-domain and frequency-domain analyses of the same data. 

For future analyses, the noise of the SQUID detection system has been reduced down to $\delta B \sim 160\,\rm{aT}\sqrt{\rm{Hz}}$ \cite{newsquid}. Further improvements come about from increased densities of He and Xe in the cell, which increases the amplitude of the transverse magnetization oscillation, in turn increasing the axion signal. These magnetometers show extreme sensitivity to these axions, allowing us to set limits past the astrophysical bounds within current experimental parameters.

\section{Atom Interferometry}
\label{sec:atom_intf}

The spin coupling of the atoms to the axion field not only induces classically intuitive effects such as Larmor precession, but also induces a phase shift as the atoms evolve due to the presence of the modified Hamiltonian. These atomic phase shifts are naturally measured using atom interferometry, where the phase shift can be measured by interfering atoms in different spin states. This type of spin interferometry has not yet been demonstrated, but there are several systems that naturally lend themselves to this type of measurement. Furthermore, the additional degree of freedom provided by measurements of different spin states allows us to construct different interferometers at once, which we can exploit to cancel off many noise sources to measure at the shot noise limit. In this section we will discuss a specific implementation of this type of atom interferometer, describe a potential strategy for removing sources of noise, and discuss the sensitivity of this type of experiment to an ultralight axion field. 

The phase shift from the axion field can be understood to lowest order as a differential phase shift between two atoms in different spin states $m_{S, 1}$ and $m_{S, 2}$. After some interrogation time $T$, this phase shift takes the form: 

\begin{equation}
\Delta\phi = (m_{S, 1} - m_{S, 2})g_{\rm{aNN}}\frac{v\sqrt{2\rho_{\rm{DM}}}}{m_{\rm{a}}}\sin{ m_{\rm{a}} T}
\end{equation}

A natural candidate for this search is the $^1S_0 - ^3P_0$ optical clock transition in alkaline-earth atoms, including Ytterbium and Strontium. This type of interferometer has recently been proposed to search for gravitational waves \cite{PhysRevLett.110.171102} and for scalar dark matter \cite{Arvanitaki:2016fyj}, so this analysis is a natural extension into pseudoscalar dark matter. The fermionic isotopes of these elements have nuclear spin that arises from a combination of neutron spin and neutron angular momentum, which causes a hyperfine splitting of both of the clock states into a manifold of different spin states, where the stretched hyperfine states are the spin-up and spin-down states of the neutron spin. Tthe phase shift can then be measured between the $^1S_0 \left(m_S = \pm \frac{1}{2}\right)$ state and the $^3P_0 \left(m_S = \mp \frac{1}{2}\right)$ state. 

\subsection{Experimental Proposal}

\subsubsection{Standard Broadband Analysis}

As described in the first section, we expect the axion dark matter to create a spurious phase shift on a spin due to the precession of the spin around the axion field. This phase shift can be measured using atom interferometry by interfering two atoms in different nuclear spin states, allowing us to probe the $g_{\rm{aNN}}$ coupling. For this analysis, we will focus on the interferometer proposed in \cite{PhysRevLett.110.171102} that is based on $^{87}$Sr, which has a a nuclear spin of $I = \frac{9}{2}$. From the nuclear shell model, we know that this means that the nucleus is in a well-defined nucleon spin state for the stretched states, $m_I = \pm \frac{9}{2}$, where we have $m_{N} = \pm \frac{1}{2}$ \cite{RevModPhys.77.427}. We will have a total spin difference of $\Delta m = 1$, which we would expect for the coupling of the axion to any fundamental fermion. 

The presence of an oscillatory phase shift naturally separates the analysis of the phase shift into two distinct regions: $m_{\rm{a}} < \frac{1}{T}$, where the interferometer will not integrate over enough time to allow a complete period of the axion field, and $m_{\rm{a}} > \frac{1}{T}$, where the phase shift will undergo multiple oscillations during the interferometer interrogation. These two regions will have qualitatively different behavior. Modern atom interferometry experiments can achieve interrogation times of $T \sim 1\,\rm{s}$, so we can estimate the maximum phase shift at the dark matter mass $m_{\rm{a}} = 2\pi\times T^{-1} \simeq 2\pi \times 1 \,\rm{Hz}$:

\begin{equation}
\Delta\phi \sim 2\times10^{-10}\left(\frac{\Delta m_S}{1}\right)\left(\frac{g_{\rm{aNN}}}{10^{-10}\;\mathrm{GeV}^{-1}}\right)\left(\frac{v}{10^{-3}}\right)\sqrt{\frac{\rho_{\rm{DM}}}{(0.04 \;\mathrm{eV})^4}}\left(\frac{2\pi \times 1 \;\mathrm{Hz}}{m_{\rm{a}}}\right)
\label{eq:phase_shift}
\end{equation}

For masses below the inverse free fall time, only a small portion of the total axion period contributes to the phase shift. We can approximate the phase shift above by $\phi \simeq \frac{v\sqrt{2\rho_{\rm{DM}}}}{m_{\rm{a}}}(m_{\rm{a}}T) = v\sqrt{2\rho_{\rm{DM}}}T$. We expect the phase shift should be independent of axion mass and should grow with larger interrogation times. For masses above the free fall time, the axion will undergo multiple periods during the interrogation, so we will only be sensitive to the amplitude of the phase shift oscillation, which scales with the axion mass as $m_{\rm{a}}^{-1}$.  We expect to quickly lose sensitivity to higher mass axions. This method also produces cusps located at $m_{\rm{a}} T = n\pi$, but in practice, these cusps can be avoided by running the interferometer with different interrogation times between runs. 

\subsubsection{High-Frequency Resonant Analysis}

We can potentially improve this experiment by taking advantage of the multiple oscillations of the axion field through an interferometric sequence that transitions between the spin states at the frequency of the axion, allowing the phase shift to accumulate. This is done by using a laser to transfer the atom population between the spin-up $m_N = +\frac{1}{2}$ and the spin-down state $m_N = -\frac{1}{2}$. If the frequency of the spin-flipping is equal to the axion mass, then we can flip the sign of the phase shift after each oscillation to allow the phase shift to build up over the coherence time of the axion. Experimentally, this spin-flipping can be accomplished in sub-millisecond time scales, allowing us to probe up to 1 kHz ($\sim 10^{-13}$ eV) axion masses. 

The integration time for the resonant experiment is set for the purposes of these estimates by the requirement that we want to probe a decade of axion masses in the same integration time as the broadband experiment ($T_{\rm{int}} \sim 1 \;\rm{year}$). This will allow us to probe a few decades of axion masses in a reasonable amount of time. The time spent in each bin is thus $t_{\mathrm{bin}} = \frac{T_{\rm{int}}}{N Q_\mathrm{eff}}$, where $N$ is the range of masses traversed ($N = 10$ for a decade of masses) and $Q_{\mathrm{eff}}$ is the effective width of each bin. This is taken to be $Q_{\mathrm{eff}} = \mathrm{min}\left(10^6, \frac{m_{\rm{a}} T}{\pi}\right)$. In practice, for the masses of interest, $\frac{m_{\rm{a}} T}{\pi} < 10^6$, so this is our limiting bin width. We have $t_{\mathrm{bin}} = \frac{\pi}{N m_{\rm{a}} T} T_{\rm{int}}$. This causes the sensitivity to fall off like $m_{\rm{a}}^{-1/2}$, improving the sensitivity to higher masses as compared to the broadband case.  However, this method is also limited by the number of pulses that the interferometer can make within the integration time before atom loss becomes significant. For realistic atom number losses, we take this number to be $\mathcal{O}(10^3)$. 

\subsection{Noise Cancellation Scheme}
%

Atom interferometry is plagued by several major noise sources, the most prominent of which for this analysis are background magnetic fields and laser phase noise. To observe an axion with a coupling of $g_{\rm{aNN}} \sim 10^{-10} \; \mathrm{GeV}^{-1}$, the background magnetic field that must be controlled at a level of $\delta B = 10^{-15}$ T, which is difficult, although manageable with magnetic shielding (further discussion of this point in Appendix \ref{sec:backgrounds}). However, the degrees of freedom provided by the different hyperfine states allow us to make multiple measurements to subtract out these noise sources. In this section we propose a method for removing these sources of noise through interferometry on multiple spin state transitions.

The phase measurement of interest between the $^1S_0 \left(m_N = \pm \frac{1}{2}\right)$ state and the $^3P_0 \left(m_N = \mp \frac{1}{2}\right)$ state is proportional to the axion signal, to laser phase noise, and to magnetic field backgrounds according to the relative Zeeman coefficient between the two states, which has a size $g_S + g_P \sim 160 \;\rm{Hz}/\rm{G}$. We want to make many simultaneous measurements to reduce these backgrounds. 

While making the axion phase measurement, we can simultaneously measure the magnetic field with the same set of atoms by using the same $^1S_0$-$^3P_0$ transition but using states of the same spin, i.e. $^1S_0 \left(m_N = \pm\frac{1}{2}\right)$ -$^3P_0 \left(m_N = \pm\frac{1}{2}\right)$. These two measurements have equal contributions from the laser phase, while having equal and opposite magnetic field phase information with a Zeeman coefficient of $g_S - g_P$. Both contributions can then be measured through symmetric and anti-symmetric combinations of these measurements, allowing us to subtract off the different contributions. The magnetic field contribution must be subtracted according to the relative Zeeman coefficients, but since $\frac{g_s + g_p}{g_s - g_p} \sim \mathcal{O}(1)$, we know that the size of the magnetic field signal is of the same order of magnitude in both measurements.

Since all measurements are shot-noise limited, we can subtract out each of the noise contributions at a sufficient level to measure the signal. With adequate magnetic shielding as discussed in Appendix \ref{sec:backgrounds}, the magnetic field noise does not limit the axion measurement for a coupling $g_{aNN} \simeq 10^{-10} \, \rm{GeV}^{-1}$, but smaller couplings will require this subtraction scheme. It is important to note that making all measurements reduces the effective shot noise of the measurement by a factor of $\sqrt{3}$ due to the reduction in the effective number of atoms, but this can be overcome by increasing the number of atoms in the interferometer sequence. We leave a dedicated experimental proposal for this noise-cancellation scheme to future study.




\subsection{Sensitivity Estimate}

\begin{figure*}
\includegraphics[width = 0.75\textwidth]{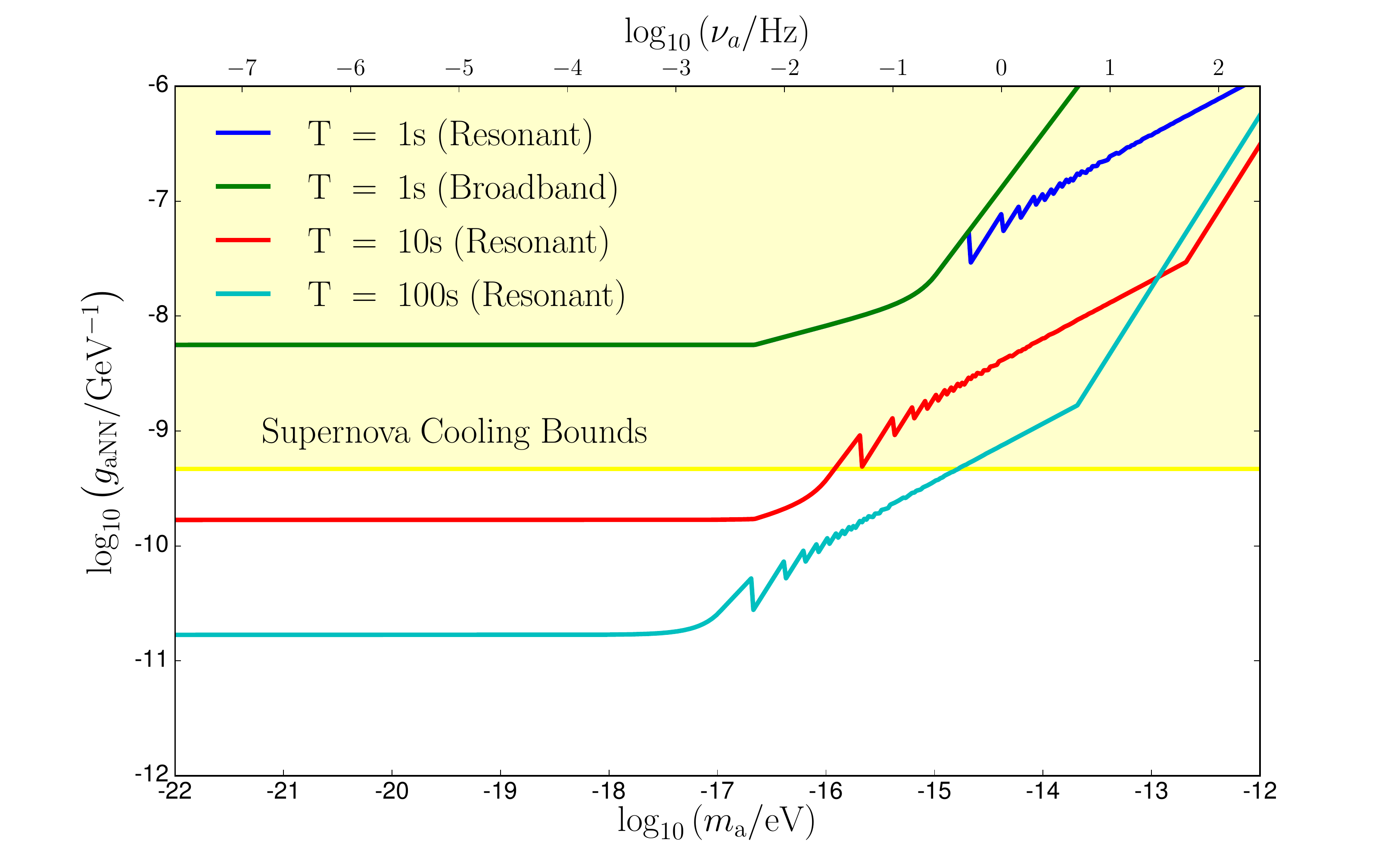}
\caption{Estimate of the exclusion sensitivity reach of the atom interferometer for the $g_{\rm{aNN}}$ coupling over an integration time of 1 year. Again, we see a flat frequency response at low frequencies due to the effective DC response of the phase shift. The signal reduces at higher frequencies in the broadband experiment as the phase shift is only sensitive to the amplitude of the oscillations. This can be improved with a resonant experiment that amplifies the phase shift at the axion frequency, which then falls off due to the amount of time spent in each frequency bin. Future experiments are primarily improved by reducing shot noise and increasing interrogation time of the experiment.}
\label{atom_int_sens_est}
\end{figure*}

With this noise-cancellation scheme in mind, we can thus calculate the sensitivity of the atom interferometer using the shot-noise limit of $\delta{\phi} = \frac{1}{\sqrt{N}}$, where $N$ is the number of atoms per second. For this analysis, we take $N = 10^8$ atoms/s, giving a shot noise of $\delta\phi = 10^{-4} \,\mathrm{rad}/\sqrt{\mathrm{Hz}}$. We find the sensitivity in Figure \ref{atom_int_sens_est}, assuming a total integration time of 1 year. We plot the sensitivity for both the resonant and broadband experiments to demonstrate their relative sensitivity for current atom interferometers with interrogation times of $T \sim 1s$. We also plot the sensitivity for resonant future atom interferometers that are in development as gravitational wave detectors. These proposals include both a ground-based atom interferometer with interrogation times up to $T = 10s$, and a space-based atom interferometer with an extremely long baseline that allows for interrogation times of up to $T = 100s$. Both of these proposals also include the possibility of using an increase shot repetition rate of up to $10 \; \mathrm{Hz}$, effectively increasing the shot noise to $\delta\phi \simeq 3\times 10^{-5} \,\rm{rad}/\sqrt{\rm{Hz}}$. These are all plotted in Figure \ref{atom_int_sens_est}, showing that this will improve the sensitivity to the axion by up to two orders of magnitude, probing past the astrophysical bounds.

We find that this experiment is particularly sensitive to axions right around the transition mass of $m_{\rm{a}} = \frac{\pi}{T}$. Below this mass, the interferometer is only sensitive to the total phase accumulated during the interrogation time. Right above the transition mass, we see that the broadband and resonant experiment have nearly the same sensitivity, but the resonant experiment quickly provides greater sensitivity despite the loss of integration time. The resonant experiment shows strong sensitivity for several decades after the transition point, at least up to the second kink where the limited number of laser pulses becomes an issue.

Further improvements include ``bouncing'' the atoms in the interferometer to increase the effective interrogation time, as well as spin squeezing to improve the signal to noise ratio towards the Heisenberg limit. Current squeezing experiments have demonstrated squeeze factors of $\sqrt{N} \sim 100$, providing large signal boosts as well as relaxing the requirement on atom number \cite{squeezing}. These squeezing techniques have not yet been demonstrated in the context of atom interferometry, but in an optimistic scenario, we could expect at least an order of magnitude improvement from squeezing.

\section{Vector Couplings}
\label{sec:vectors}

This oscillating spin-coupling is not unique to the axion, as it can arise from any operator that couples an external classical field to fermions. In particular, this same coupling can arise from the coupling of a spin-1 vector, commonly known as the ``hidden photon''. These couplings take similar forms to the axion couplings described above, and thus the related plots will look identical to those in the previous section, as we show below. In particular, we will focus on the dimension 5 electric and magnetic dipole moment operators due to their similarity to the axion operator, and on the kinetic mixing term induced at one loop by this operator. 


%

First we consider the magnetic dipole moment operator, which couples the hidden photon field to the spins of electrons or nucleons. Following the example of the magnetic dipole moment operator for photons, we can write this coupling in the Lagrangian as
\begin{equation}
\mathcal{L} \supset g_{A\psi\psi}F^{\prime}_{\mu\nu}\bar{\psi}\sigma^{\mu\nu}\psi 
\end{equation}
where $F^{\prime}_{\mu\nu} = \partial_{\mu}A^{\prime}_{\nu} - \partial_{\nu}A^{\prime}_{\mu}$, i.e. the field strength of the hidden photon field, and $\psi$ stands for either an electron or nucleon. Under the same assumption as in the axion case that the hidden photon field can be modeled as a classical wave and following the standard formulation of the electromagnetic field strength tensor, we can estimate the size of the hidden electric and magnetic fields as $E^{\prime} \sim B^{\prime}/v \sim \sqrt{2\rho_{\rm{DM}}}$. We can estimate the direction of the electric field as $\vec{E^{\prime}} \sim \partial_0 \vec{A}^{\prime}$ since the spatial gradient is small. Thus, $\vec{E}^{\prime}$ points in the same direction as $\vec{A}^{\prime}$, which we define as $\hat{n}$. We can then estimate the direction of the magnetic field as $\hat{B}^{\prime} \sim \hat{v} \times \hat{n}$. This then allows to approximate this coupling in terms of a non-relativistic Hamiltonian: 
\begin{equation}
\label{eqn: hidden magnetic dipole}
H \simeq g_{A\psi\psi}\vec{B'}\cdot\vec{\sigma}_{\psi} \sim g_{A\psi\psi}\sqrt{2\rho_{\rm{DM}}}(\vec{v}\times\hat{n})\cdot\vec{\sigma}_{\psi}
\end{equation} 

From equation \eqref{eqn: hidden magnetic dipole} it is clear that the sensitivities of our experiments for the coupling $g_{A\psi\psi}$ will be identical to those for the axion spin couplings we have already considered. The primary difference is the current bounds on this coupling that exist from astrophysical sources. We can estimate similar astrophysics bounds as for the axion by considering the influence of hidden photons on the cooling rates of stars. The processes that contribute to this cooling rate are identical for this particular axion and hidden photon coupling. In fact, the cross section for the hidden photon processes are exactly twice that of the axion processes due to the multiple polarization states of the hidden photon \cite{PhysRevLett.94.151802}. The bounds shown in the plots above are thus strengthened only by a factor of $\sqrt{2}$. 
Thus the plots for hidden photon dark matter should look very similar to those for the axion, Figures \ref{pend_sens_est}, \ref{fig:mag_sensitivity}, and \ref{atom_int_sens_est} above.

Much like the magnetic dipole moment, there is also an associated electric dipole moment term for the hidden photon. For a normal photon, the electric dipole moment coupling to spin vanishes due to CP symmetry. However, because the hidden photon could also be CP violating, we can just as easily have an electric dipole term for the hidden photon. Again following the form of the electric dipole moment operator for photons and using our same assumptions for the hidden photon as above, we model this interaction as:
\begin{equation}
\begin{gathered}
\mathcal{L} \supset g_{A\psi\psi}F^{\prime}_{\mu\nu}\bar{\psi}\gamma_5\sigma^{\mu\nu}\psi \\
\Rightarrow H \simeq g_{A\psi\psi}\vec{E'}\cdot\vec{\sigma}_{\psi} \sim g_{A\psi\psi}\sqrt{2\rho_{\rm{DM}}}\hat{n}\cdot\vec{\sigma}_{\psi}
\end{gathered}
\end{equation}
For a hidden photon, since we know that $E^{\prime} \sim \sqrt{2\rho_{\rm{DM}}}$, the size of the signal is no longer suppressed by the velocity of the dark matter. We would expect the signal from a hidden photon electric dipole coupling to be $10^3$ times bigger than that of the magnetic dipole coupling. All of the sensitivity curves shown above would then be pushed down by 3 orders of magnitude.  However the astrophysical bounds are the same as those for the magnetic dipole coupling, giving our experiments a much larger reach beyond current constraints in this electric dipole coupling. 
%
%

In addition to the direct dipole moment measurements, this operator induces kinetic mixing effects between the photon and the hidden photon through loop diagrams.   This coupling can cause anomalous oscillating electromagnetic effects that have been the subject of other experiments \cite{Graham:2014sha, Chaudhuri:2014dla}.
This kinetic mixing coupling term can be diagonalized into the form 
\begin{equation}
\label{eqn: kinetic mixing}
\mathcal{L} \supset - eJ^{\mu}_{\rm{EM}} (A_{\mu} + \varepsilon A^{\prime}_{\mu})
\end{equation}
where $A_{\mu}$ is the standard photon operator, $A^{\prime}_{\mu}$ is the hidden photon operator, and $J^{\mu}_{\rm{EM}}$ is the electromagnetic current. The photon couples to the current through the standard coupling $e$, while the coupling to the hidden photon is suppressed by an additional factor of $\varepsilon$, which quantifies the degree of kinetic mixing between the hidden photon and the standard model.
There is a minimum natural size to this coupling (without tuning):
\begin{equation}
\varepsilon \sim 10^{-10}\left(\frac{g_{\rm{aNN}}}{10^{-10}\;\mathrm{GeV^{-1}}}\right)\log\left(\frac{\Lambda}{m_{A^{\prime}}}\right)
\end{equation}
where $m_{A^{\prime}}$ is the mass of the hidden photon and $\Lambda$ is the scale of the theory cutoff. 
The largest possible hierarchy of scales would be for $\Lambda \sim M_{\rm{pl}}$ and $m_{A^{\prime}} \sim 10^{-22} \; \mathrm{eV}$, which gives $\log{\frac{\Lambda}{m_{A^{\prime}}}} \sim 100$. 
If we use this formula to translate the current bounds on $\varepsilon$ for our mass range, they are substantially weaker than the astrophysical bounds we have already plotted and thus irrelevant.

It is also possible that these experiments could directly bound a hidden photon that has only the kinetic mixing (resulting in equation \eqref{eqn: kinetic mixing}).  In this case, the kinetic mixing can induce a hidden magnetic dipole moment down from the standard model value for each fermion by a factor of $\varepsilon$.  Inside a shield, the hidden photon dark matter looks like an effective magnetic field \cite{Chaudhuri:2014dla}.  Thus all co-magnetometry is not useful and we would simply have to rely on magnetic shielding to distinguish this from background magnetic noise, which is nontrivial.  To compute the sensitivity of any of our experiments to this kinetic mixing of the hidden photon we can use the following formula to convert sensitivities on $\coupling$ from above to sensitivities on $\varepsilon$:
\begin{equation}
\varepsilon \sim \frac{m_\psi}{\alpha} v \, m_{A'} \, L_\text{exp} \, \coupling
\end{equation}
where $L_\text{exp}$ is the rough size of the magnetically shielded region.  Using this estimate, the most optimistic future experiments will get roughly to the astrophysical bound from heating of the interstellar medium \cite{Dubovsky:2015cca}.

\section{Conclusions}
\label{sec:conclusions}


Ultralight, axion-like dark matter has recently attracted a great deal of attention.
While several axion dark matter experiments exist, none are sensitive to masses below $10^{-13}$ eV ($\sim 10^2$ Hz).
This range is theoretically well-motivated, including for example the possibility of ``fuzzy dark matter" which may resolve problems in structure formation.
In this paper, we have described how the low-mass frontier of axion dark matter produces effects, including a time-oscillating torque and a variation in spin precession frequency, that can be probed using existing experimental techniques. The frequency and direction of these effects are defined by fundamental physics, allowing these experiments to better distinguish this dark matter signal from normal laboratory backgrounds. 

We have found three possible, high-precision experimental techniques to search for axion dark matter in this mass range.
Torsion balances, atomic magnetometers, and atom interferometers can all be used detect the effects of axion dark matter on the spins of electrons or nuclei.
Torsion balances and atomic magnetometers have already placed constraints on static Lorentz invariance violation, allowing us to reinterpret these bounds as bounds on axion dark matter.
The potential sensitivity of these techniques is shown in Figures \ref{pend_sens_est}, \ref{fig:mag_sensitivity}, and \ref{atom_int_sens_est}. 
These experiments are rapidly improving in sensitivity, and future iterations will soon be able to push beyond the astrophysical bounds into uncharted territory for axion dark matter. 

\section*{Acknowledgements}
We acknowledge useful conversations with Dmitry~Budker, Jason~Hogan, Silvia Knappe-Gr{\"u}neberg, David Marsh, Yevgeny Stadnik, Alex~Sushkov, and Victor Flambaum.
TW is supported by the Department of Defense (DoD) through the National Defense Science and Engineering Graduate Fellowship (NDSEG) Program.  PWG acknowledges the support of NSF grant PHY-1720397 and the DOE Early Career Award DE-SC0012012.  DEK acknowledges the support of NSF grant PHY-1214000. SR was supported in part by the NSF under grants PHY-1417295 and PHY-1507160, the Alfred P. Sloan Foundation grant FG-2016-6193, the Simons Foundation Award 378243. WT was supported by the Alexander von Humboldt Foundation and the German Federal Ministry for Education and Research. This work was supported in part by the Heising-Simons Foundation grants 2015-037 and 2015-038 and the W.M. Keck Foundation.

\appendix
\section{Magnetic Field Backgrounds}
\label{sec:backgrounds}

The primary background in all of these experiments is due to magnetic fields, which couple to spin according to the magnetic moment of the particular fermion. Roughly, we can estimate the necessary magnetic field noise suppression by calculating the size of the minimum magnetic field necessary to mask the axion signal, and requiring that the total measured noise within the signal window is smaller than this limit. In other words, we require

\begin{equation}
\delta B_{\mathrm{noise}} = \sqrt{P_B^2 \times \mathrm{max}\left(\frac{m_{\rm{a}}}{Q}, \frac{\pi}{T_2}\right)} \leq \delta B_{\mathrm{sig}} = \frac{g_{a\psi\psi}\vec{\nabla}a \cdot \vec{\sigma}_{\psi}}{\mu_{\mathrm{\psi}}}
\end{equation}

\begin{figure*}
\includegraphics[width = 0.75\textwidth]{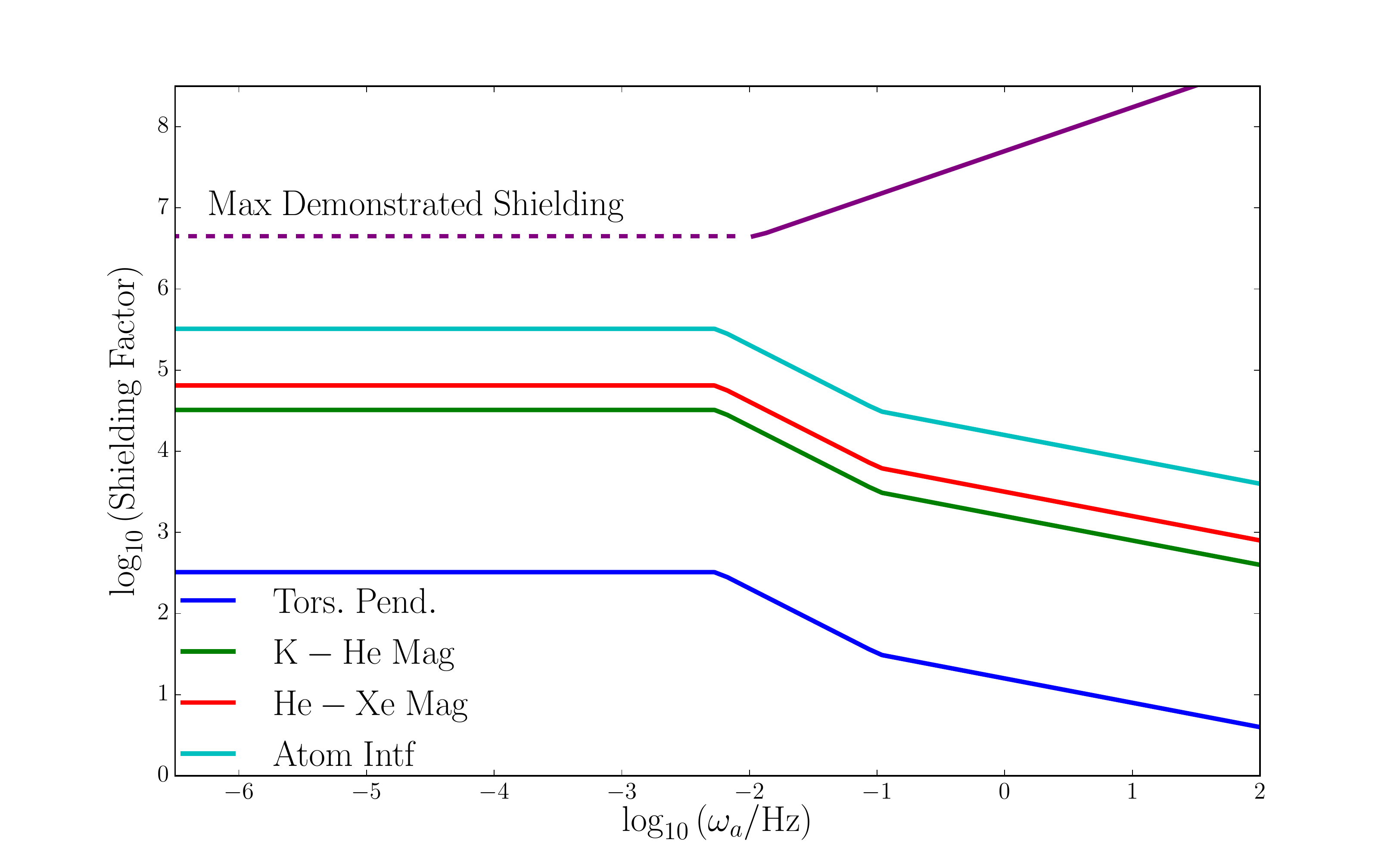}
\caption{This figure shows the magnetic shielding required for each experiment to measure a signal of coupling strength $g_{A\psi\psi} \sim 10^{-10} \;\mathrm{GeV}^{-1}$, assuming a typical background magnetic field noise spectrum as measured in \cite{Klein_the8-layered, mag_room}. Existing reduction of the noise due to co-magnetometry effects are specific to each experiment and are discussed in the text. This is compared to the maximum demonstrated shielding in purple as measured for high frequencies (solid line) and projected down to lower frequencies (dashed line) based on typical static shielding factors. Since all of experiments mentioned here have a required shielding factor below the maximum level, this shows the magnetic field noise can be successfully suppressed to measure this coupling strength. Further coupling measurements will require more stringent background contro, which can be scaled from the plot. These are only rough approximations to the magnetic shielding, further analysis will be required for each dedicated experiment to better understand how to control the magnetic field backgrounds.}
\label{fig:mag_noise}
\end{figure*}

Knowing the size of the signals as described above, we can calculate the required magnetic field noise suppression (and thus the level of magnetic shielding required) for each experiment. We can invert the above expression to solve for the noise spectrum $P_B$ and compare this to measured background levels \cite{Klein_the8-layered, mag_room} for a fixed value of the coupling, which we take to be $g_{\rm{aNN}} = 10^{-10} \; \mathrm{GeV}^{-1}$. For the background magnetic field, we assume that the noise reaches a DC limit at $10^{-7} \;\rm{T}$, which has been demonstrated by many experiments. To calculate the bandwidth of the experiment, we use a total signal time of $T = \min(\frac{Q}{m_{\rm{a}}}, T_{\rm{int}})$, which leads to the kink seen at $100 \; \rm{mHz}$. 

For the experiments in question, we can calculate the size of the magnetic field effect by using the magnetic moment of the particular experiment (i.e. the magnetic moment of the particle in question) and by accounting for co-magnetometry effects that reduce the overall magnetic field background noise. For the torsion pendulum, the co-magnetometry effect is already taken into account into the reduction of the overall magnetic moment of the torsion pendulum down to $10^{-5} \;\mathrm{J/G}$ \cite{PhysRevD.78.092006}. For the K-He magnetometer, the co-magnetometry effect has been measured as having a further suppression of the noise of $10^{4}$ for the electron magnetic moment \cite{PhysRevLett.105.151604, Brown:2011}. For the He-Xe magnetometer, the co-magnetometry is difficult to quantify simply, as it is dependent on the time-dependent drifts in the magnetic field. However, for simplicity, we assume a constant conservative shielding factor of $10^2$ on the nuclear magnetic moment as demonstrated by the measurement in \cite{PhysRevD.82.111901}. The co-magnetometry measurement for atom interferometry as described in Section \ref{sec:atom_intf} is only a theoretical measurement and has not been characterized, so we can not accurately model it in this analysis. However, we can see that even in the absence of co-magnetometry, the shielding requirements for atom interferometry are still achievable with current technology.

We can further reduce the size of the magnetic fields using external magnetic shielding. Using the measurement of background magnetic fields \cite{Klein_the8-layered, mag_room}, and accounting for co-magnetometry effects above, we can estimate the shielding factor required to measure a signal at the $g_{\rm{aNN}} = 10^{-10} \;\rm{GeV}^{-1}$ level. These shielding factors are shown in Figure \ref{fig:mag_noise} compared to a realistic shielding factor that has been demonstrated by state-of-the-art $\mu$-metal shields. For this comparison, we plot measured shielding values for various frequencies (shown by the solid line) and then project them down to lower frequencies (shown by the dashed line) \cite{Klein_the8-layered}. 

As shown in the figure, all experiments have a required shielding factor below realistic levels, indicating that all of the mentioned experiments can be successfully shielded to measure a coupling of $g_{\rm{aNN}} = 10^{-10} \; \rm{GeV}^{-1}$. Measurements of smaller coupling will require more shielding, and these levels can be easily understood from scaling the lines in the figure.

%

\bibliography{spin_torque_prd_resubmit}

\end{document}